\documentclass[fleqn,usenatbib]{mnras}

\usepackage{newtxtext,newtxmath}

\usepackage[T1]{fontenc}

\DeclareRobustCommand{\VAN}[3]{#2}
\let\VANthebibliography\thebibliography
\def\thebibliography{\DeclareRobustCommand{\VAN}[3]{##3}\VANthebibliography}

\usepackage{graphicx}	
\usepackage{amsmath}	
\usepackage{longtable}

\title[Structure in the break]{GRB jet structure and the jet break}

\author[G. P. Lamb et al.]{Gavin P. Lamb$^{1}$\thanks{E-mail: \href{mailto:gpl6@leicester.ac.uk}{gpl6@leicester.ac.uk}},
{D. Alexander Kann$^{2}$,}
{Joseph John Fernández$^{3}$,}
{Ilya Mandel$^{4,~5,~6}$,}
{Andrew J. Levan$^{7}$,}
\newauthor
{\& Nial R. Tanvir$^{1}$.}
\\
$^{1}$School of Physics and Astronomy, University of Leicester, University Road, Leicester, LE1 7RH, UK\\
$^{2}$IAA-CSIC, Glorieta de la Astronom\'{i}a, s/n, 18008 Granada, Spain\\
$^{3}$Astrophysics Research Institute, Liverpool John Moores University, IC2, Liverpool Science Park, 146 Brownlow Hill, Liverpool L3 5RF, UK\\
$^{4}$Monash Centre for Astrophysics, School of Physics and Astronomy, Monash University, Clayton, Victoria 3800, Australia\\
$^{5}$OzGrav, Australian Research Council Centre of Excellence for Gravitational Wave Discovery, Australia\\
$^{6}$Institute of Gravitational Wave Astronomy and School of Physics and Astronomy, University of Birmingham, Birmingham, B15 2TT, United Kingdom\\
$^{7}$Department of Astrophysics, Radboud University, 6525 AJ Nijmegen, The Netherlands
}

\date{Accepted XXX. Received YYY; in original form ZZZ}

\pubyear{2021}

\begin{document}
\label{firstpage}
\pagerange{\pageref{firstpage}--\pageref{lastpage}}
\maketitle

\begin{abstract}
We investigate the shape of the jet break in within-beam gamma-ray burst (GRB) optical afterglows for various lateral jet structure profiles. We consider cases with and without lateral spreading and a range of inclinations within the jet core half-opening angle, $\theta_c$. We fit model and observed afterglow lightcurves with a smoothly-broken power-law function with a free-parameter $\kappa$ that describes the sharpness of the break. We find that the jet break is sharper ($\kappa$ is greater) when lateral spreading is included than in the absence of lateral spreading. For profiles with a sharp-edged core, the sharpness parameter has a broad range of $0.1\lesssim\kappa\lesssim4.6$, whereas profiles with a smooth-edged core have a narrower range of $0.1\lesssim\kappa\lesssim2.2$ when models both with and without lateral spreading are included. For sharp-edged jets, the jet break sharpness depends strongly on the inclination of the system within $\theta_c$, whereas for smooth-edged jets, $\kappa$ is more strongly dependent on the size of $\theta_c$. Using a sample of 20 GRBs we find nine candidate smooth-edged jet structures and eight candidate sharp-edged jet structures, while the remaining three are consistent with either. The shape of the jet break, as measured by the sharpness parameter $\kappa$, can be used as an initial check for the presence of lateral structure in within-beam GRBs where the afterglow is well-sampled at and around the jet-break time.
\end{abstract}




\section{Introduction}

The achromatic breaks observed in the lightcurves of some gamma-ray burst (GRB) afterglows have been used to argue that these sources are jet-like in nature \citep{rhoads1997}.
The observed time of the afterglow  jet break after the prompt GRB contains information about the angular size, $\theta_j$ of the outflows that produce these transients \citep[e.g.][]{halpern1999, sari1999, jaunsen2001, panaitescu2002} and this knowledge of a finite angular size relaxes the energy requirements needed to produce cosmological GRBs when assuming isotropic emission \citep{kulkarni1999}.
The temporal index $\alpha$ of the lightcurve before a jet break, and the steeper decline $\alpha_2$ after it, provide information about the ambient medium density profile and the spectral regime of the emission \citep{granot2002} and constrain energy injection into the outflow  \citep{zhang2006}.

The change in the temporal index, $\Delta\alpha$, across the jet break can be used to indicate the degree of lateral spreading perpendicular to the outflow radial motion.
For a jet without any lateral spreading, the loss of flux due to the edge of the jet becoming visible as the Lorentz factor decreases, $\Gamma\leq\theta_j^{-1}$, yields $\Delta\alpha=3/4$ for a uniform ambient medium, or $\Delta\alpha=1$ for a wind-like medium \citep{rhoads1999, sari1999, gao2013}.
Where lateral spreading is included,  the post-break decline index is $\alpha\leq -p$, where $p$ is the electron energy spectral slope \citep[e.g.][]{rhoads1999, sari1999, zhang2009}.

The shape of the jet break -- how rapidly the lightcurve behaviour changes from the pre- to the post-break temporal behaviour -- depends on the angular size of the jet and the inclination of the line-of-sight within the jet opening angle, $\theta_j$.
For a jet with sharp edges, i.e. the energy declines rapidly or goes to zero at angles beyond $\theta_j$, the near edge of the jet will be viewed before the far edge for an observer whose line-of-sight is between the jet edge and the central axis. This results in a longer transition from pre- to post-jet-break behaviour, when compared with an observer that is either aligned with the central axis or the jet edge, and reduces the sharpness of the jet-break shape as measured via a smoothly broken power-law function \citep{vaneerten2013}.

Whilst looking at the time evolution of the temporal index for both sharp-edged and smooth-edged jets\footnote{A smooth-edged jet is described by a jet structure with an energy/velocity profile that varies with lateral angle, e.g. a Gaussian or power-law functional form.} at various inclinations, in a study focused on reverse-shock emission, \cite{lamb2019} showed that the temporal index, $\alpha$, for an on-axis observer exhibits different behaviour through the jet break for different jet structure models.
This suggests that the shape of the afterglow lightcurve through the jet break for on-beam (within the jet core or opening angle) cosmological GRBs could be used as an indication of lateral jet structure \citep[see also][]{rossi2004}.

In \S \ref{sec:2} we describe the methods used to generate model lightcurves and how we measure the shape of the jet break.
In \S \ref{sec:3} we present our results including the best-fit measure of the jet break shape for a sample of GRB afterglows that exhibit an achromatic jet break.  We discuss our findings in \S \ref{sec:4} and conclude in \S \ref{sec:5}.

\section{Method}\label{sec:2}

We produce afterglow lightcurves for a variety of jet structure models following the method in \cite{lamb2017, lamb2018}.
The instantaneous Lorentz factor, $\Gamma$, and swept-up mass, $M$, of a relativistic blastwave can be found by considering energy and momentum conservation \citep{peer2012, nava2013, lamb2018},
\begin{equation}
    \frac{{\rm }d \Gamma}{{\rm d} \log_{10} M} = \frac{M \ln{(10)}\left[\hat\gamma(\Gamma^2-1)-(\hat\gamma-1)\Gamma\beta^2\right]}{E/(\Gamma_0c^2)+M\left[2\hat\gamma\Gamma-(\hat\gamma-1)(1+\Gamma^{-2})\right]},
    \label{dynamics}
\end{equation}
where, $\Gamma_0$ is the maximum initial Lorentz factor, $E$ is the blastwave kinetic energy, $c$ the speed of light, $\beta = (1-\Gamma^{-2})^{1/2}$ the velocity as a fraction of the speed of light, and $\hat\gamma$ is the adiabatic index (found using the expressions given in \citealt{peer2012}).

For a conical outflow, the kinetic energy and the swept-up mass is a fixed fraction of that used in equation \ref{dynamics} for a spherically symmetric system.
The shock-heated material within the blastwave will have a significant sound speed, $c_s$, and for a conical outflow with `edges', lateral spreading will occur.
For an ideal relativistic plasma, the sound-speed limit is $c_s = c/\sqrt{3}$, and the instantaneous sound speed can be found by assuming an ideal equation-of-state, $P\propto\rho^{\hat\gamma}$, where $P$ is the pressure and $\rho$ is the density, and $c_s = (\hat\gamma P/\rho)^{1/2}$.
Using the strong relativistic shock conditions, with $P =(\hat\gamma-1) (\Gamma-1)\rho_0 c^2$, and $\rho = \rho_0 +\hat\gamma P /(c^2[\hat\gamma-1])$, the sound speed in terms of the Lorentz factor and adiabatic index is then
\begin{equation}
    c_s = \left(\frac{c^2\hat\gamma(\hat\gamma-1)(\Gamma-1)}{1+\hat\gamma(\Gamma-1)} \right)^{1/2}.
    \label{cs}
\end{equation}

As the jet spreads laterally, the half-opening angle of the outflow, $\theta_j$, will increase.
By considering an element travelling perpendicular to the outflow direction at the sound speed, and an element travelling parallel to the outflow direction at the velocity $\beta c$, the change in the initial opening angle due to lateral spreading is
\begin{equation}
    \Delta\theta \approx \tan^{-1}\left(\frac{c_s}{\Gamma\beta c}\right),
    \label{dtheta}
\end{equation}
where the sound speed, $c_s$, is defined in the co-moving fluid frame.

For an outflow where the effects of lateral spreading are included, the changing opening-angle will affect the estimate of the radial distance, $R$, that the blastwave has travelled \citep{granot2012}.
The outflow sweeps up matter and compresses it into a thin shell at the head of the jet \citep[e.g.][]{kobayashi1999}.
We approximate this thin shell of compressed matter as a surface with area, $A = \Omega R^2$, where $\Omega$ is the solid angle of the outflow. 
As mass and energy must be conserved, the area is proportional to the instantaneous swept-up mass, $M$, and for a given area and a uniform medium, the radius, $R \propto \Omega^{-1/2}$.

Practically, the radius for a spherical, or a non-laterally spreading conical segment, can be found using the swept-up mass and the ambient particle number density as, $R_i^3= 3 M_i/(\Omega n m_p)$, where $M_i$ is the mass of swept-up material for step $i$ of the solution to equation \ref{dynamics} (we solve this using a 4th order Runge-Kutta), $n$ is the ambient number density, and $m_p$ the mass of a proton.
Equation \ref{dynamics} assumes energy conservation, therefore a blastwave with a fixed initial energy has a swept-up mass at each step that does not depend on the details of lateral spreading.

The resultant solid angle for a uniform conical jet with lateral spreading and initial half-opening angle, $\theta_j$, is then $\Omega_i=2\pi(1-\cos(\theta_j+\Delta\theta_i))$, thus, where the change in radius for the blastwave is calculated from the swept-up mass at each step in the solution to equation \ref{dynamics}, the effects of lateral spreading can be included by scaling the change in radius for a spherical blastwave by the ratio, $(\Omega_0/\Omega_i)^{1/2}$, where $\Omega_0$ is the initial solid angle.

We split the jet into multiple small sections with angular size, $\theta\ll\theta_j$, and sum across the surface to get the afterglow lightcurve, effectively integrating across the equal-arrival-time surface \citep[e.g.][]{lamb2017}.
The initial angular size of each section is chosen so that $\theta\ll\Gamma_0^{-1}$, and when calculating the ratio, $\Omega_0/\Omega_i$, we set $\theta=0$ to avoid any resolution dependence.
The instantaneous radial distance at each step, $i$, for each component of the blastwave is then found with
\begin{equation}
    R_i = R_{i-1} + \left[\left(\frac{\Omega_0}{\Omega_i}\right)^{1/2}\frac{3}{4\pi n m_p}\right]^{1/3} \left(M_{i}^{1/3}-M_{i-1}^{1/3}\right),
\end{equation}
where $M$ is the swept-up mass for a spherical volume (for a non-spreading jet or spherical blastwave, the ratio $\Omega_0/\Omega_i=1$).
Note that using the ratio $\Delta\theta_0/\Delta\theta_i\sim(\Omega_0/\Omega_i)^{1/2}$, from equation \ref{dtheta} gives an identical approximation whilst the blastwave is still relativistic\footnote{The afterglow lightcurves for a `top-hat' jet structure are generated using this method and tested for consistency with the on- and off-axis lightcurves of \texttt{BoxFit} \citep{vaneerten2012}. Afterglow peak and break times plus the very late-time evolution, including the counter-jet emergence and peak time/flux, are recreated.}, and where $\theta\rightarrow0$ for each component.

\begin{figure}
    \centering
    \includegraphics[width=\columnwidth]{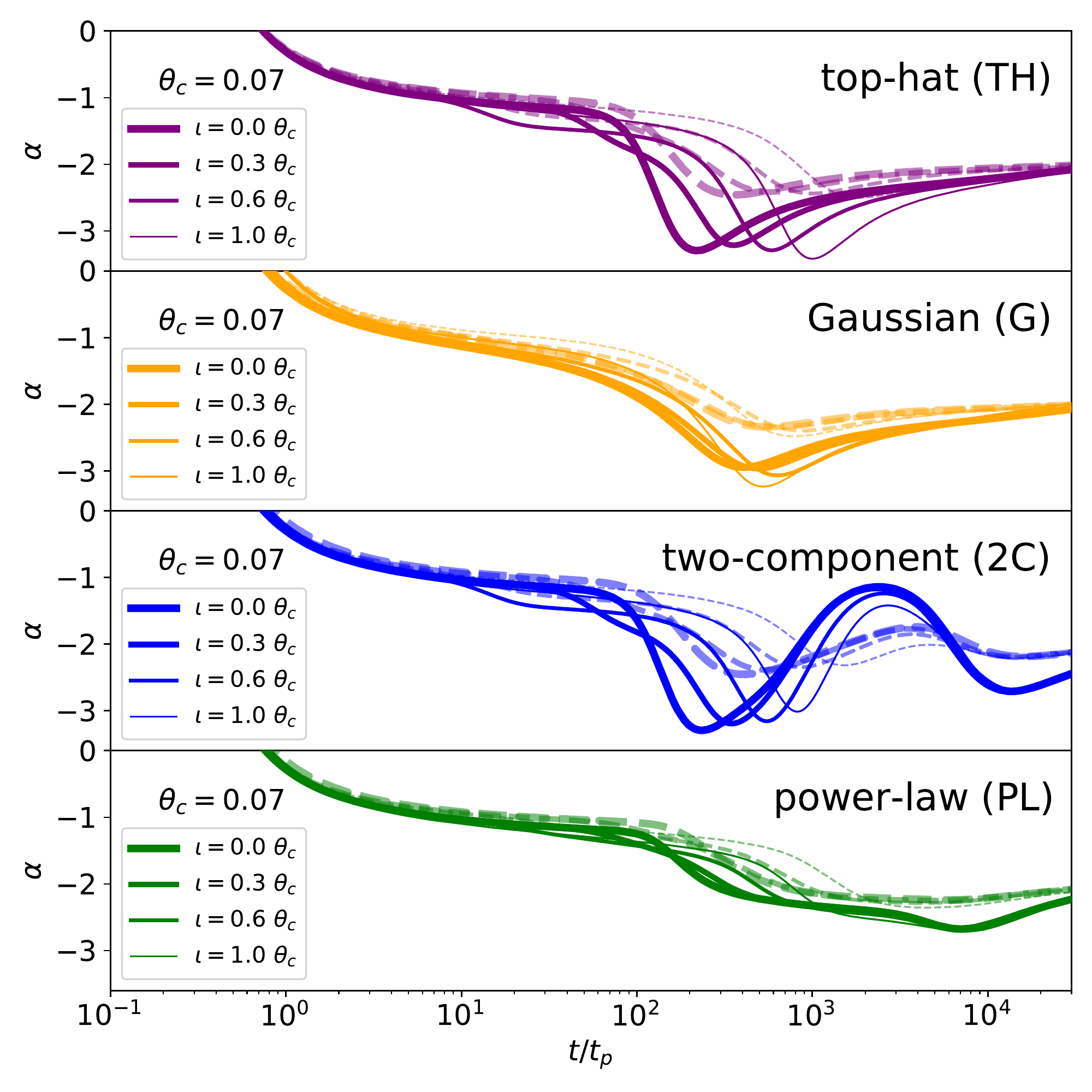}
    \caption{The evolution of the temporal index $\alpha$ with observer time in units of the afterglow peak time, $t/t_p$, for the four jet profile models with fixed $\theta_c$ and varying orientation within the core angle: $\iota = [0.0,0.3,0.6,1.0]~\theta_c$.
    Top to bottom: TH (purple), G (orange), 2C (blue), and PL (green) jet structure profiles. Solid lines show the evolution of $\alpha$ for a laterally spreading outflow, whereas the dashed lines do not include spreading. The line thickness indicates the inclination within the jet core angle.}
    \label{fig:new1}
\end{figure}

By assuming four different jet structure profiles\footnote{The jet structure profiles used are:
`top-hat' (TH), a uniform jet defined by the opening angle, $\theta_j\equiv\theta_c$, with energy $E_c$ and $\Gamma_0$;
Gaussian (G) structured jet with the lateral profile $E=E_c e^{-(\theta^2/\theta_c^2)}$ and $\Gamma= 1+(\Gamma_c-1) e^{-(\theta^2/2\theta_c^2)}$;
a two-component (2C) jet defined as a `top-hat' core surrounded by a sheath with $\Gamma=5$ and $0.1 E_c$;
and a power-law (PL) jet, a `top-hat' core with energy and Lorentz factor that declines as $\propto(\theta/\theta_c)^{-2}$ where $\theta>\theta_c$.}, as in \cite{lamb2017,lamb2019}, we generate on-axis afterglow lightcurves with and without lateral spreading to check for differences between the models. The fiducial parameters are: the isotropic equivalent energy for a point on the central axis of the jet, $E_c=10^{52}$ erg; the bulk Lorentz factor at the central jet axis, $\Gamma_0=100$; the microphysical parameters, $\varepsilon_e = \sqrt{\varepsilon_B} = 0.1$; the electron distribution index, $p=2.5$; and a uniform ambient density, $n=10^{-3}$\,cm$^{-3}$; a jet core opening angle of $\theta_c = 0.07$\,rad and a maximum jet extent of $\theta_j =0.35$\,rad where the energy/velocity is not zero at angles wider than $\theta_c$; note for TH jets, $\theta_c = \theta_j = 0.07$.
From these afterglow lightcurves, the temporal index of the afterglow with observer time is calculated from each time step using,
\begin{equation}
    \alpha = \frac{{\rm d}\log(F)}{{\rm d}\log(t_{\rm obs})},
    \label{alpha}
\end{equation}
where $F$ is the flux and $t_{\rm obs}$ the observed time.
We use Richardson's extrapolation to find $\alpha$ at each observer time step from the model afterglows.
This temporal index, $\alpha$, with the observer time relative to the optical frequency afterglow peak time, $t_p$, which is typically the deceleration time for outflows with our fiducial parameters at optical frequencies, is shown in Figure \ref{fig:new1}.

The change in the temporal index for observers at various inclinations, $\iota$, within the core angle, $\theta_c$, can be seen with thinner lines indicating a higher $\iota$.

For observers that are midway between the jet central axis and the jet edge, an earlier quasi-break can be seen for the jet structures described by a uniform core and the effect most pronounced for the sharp-edged jet structures, e.g. TH and 2C, and where lateral spreading is included. 

This quasi-break is due to the near edge of the jet becoming visible, and the main jet-break when the furthest edge becomes visible -- thus, the full jet-break is delayed when compared to that of an observer along the jet central axis.
For the TH jet model with spreading, the minimum value for $\alpha$ and its temporal behaviour is qualitatively the same as that described in \cite{vaneerten2013} via hydrodynamic simulations.

We see a similar change in minimum $\alpha$ as the jet opening angle of the system is increased and a minimum $\alpha$ that is marginally lower due to the details of the spreading -- where the sound speed expansion that is used in our model can be considered a physical upper limit.

\begin{figure*}
    \includegraphics[width=\textwidth]{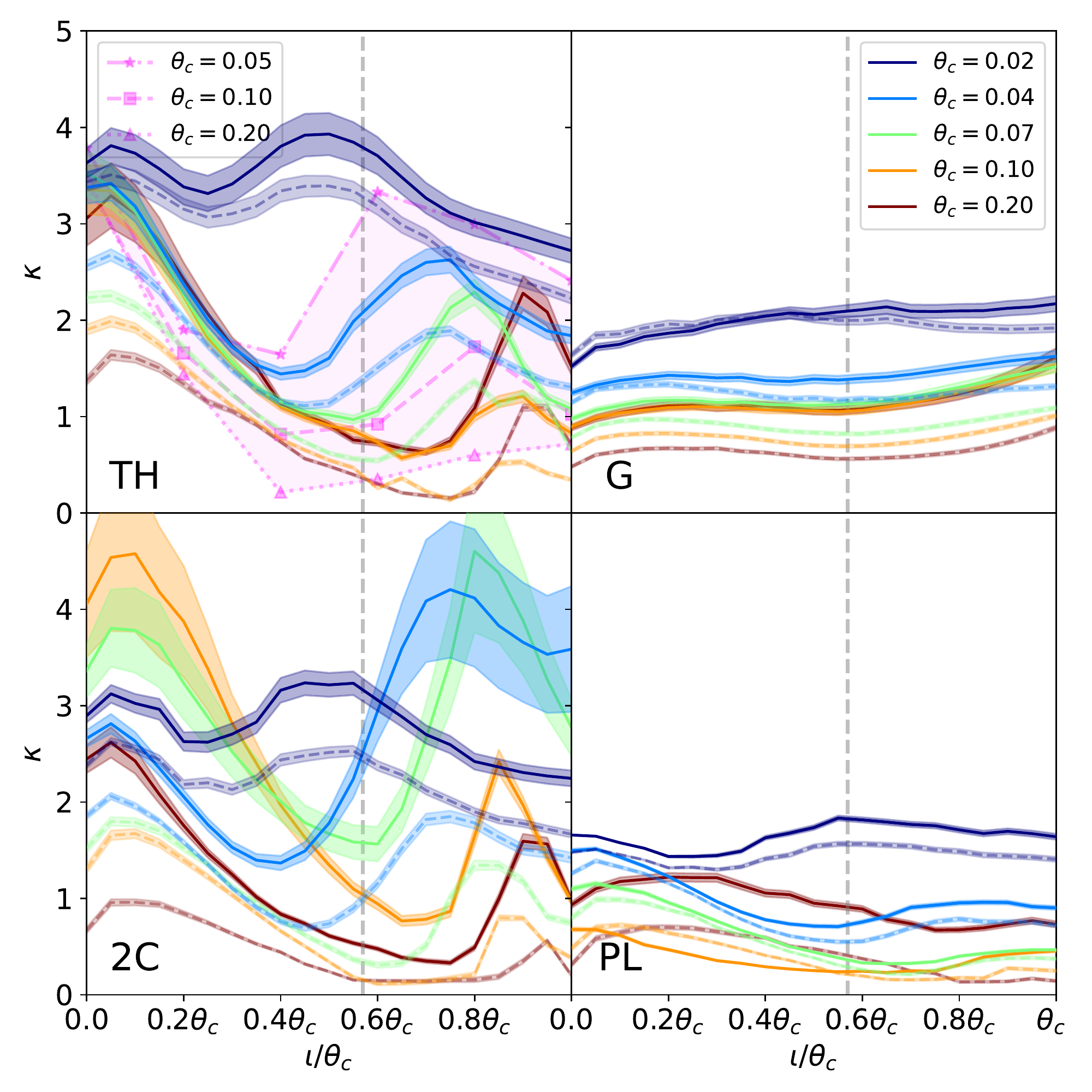}
    \caption{The $\kappa$ value vs. inclination within the jet core from fits of equation \ref{eq:sharp} to the model lightcurves at the jet-break time. The shaded regions indicate the $1\sigma$ limits for the fit for the spreading/non-spreading cases with varying core size. Solid lines indicate the central value for the maximally spreading case, while dashed lines indicate the central value for the non-spreading case. Jet structures are top left --  top-hat, top right -- Gaussian, bottom left -- two-component, and bottom right -- power-law. The pink-shaded region and lines in the top left panel (TH) show the values of $\kappa$ for top-hat jet afterglows with $p=2.5$ from \protect\cite{vaneerten2013}. The vertical grey dashed line indicates an $\iota=0.57\theta_c$, the typical inclination within the core angle for GRBs \protect\citep{ryan2015}.}
    \label{fig:new2}
\end{figure*}

\begin{figure}
    \centering
    \includegraphics[width=\columnwidth]{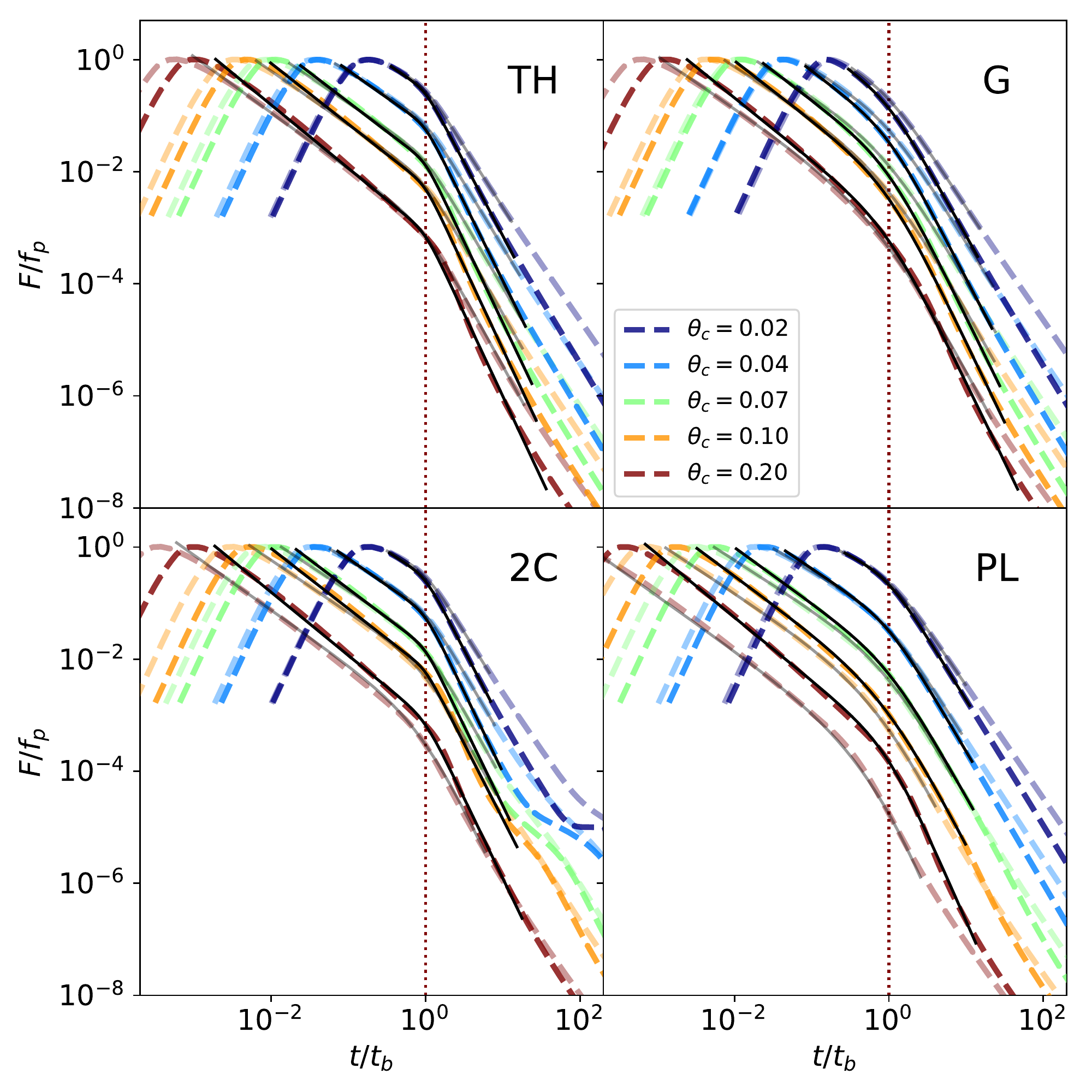}
    \caption{The on-axis, $\iota=0$, model lightcurve flux (dashed lines) as a fraction of the peak flux $f_p$, with the time normalised by the best-fit jet-break time, $t_b$, for each jet structure.  The corresponding best fits to equation \ref{eq:sharp} are shown in black/grey lines. The bolder coloured lines show the case where lateral spreading is included. The vertical dotted line shows $t=t_b$, the jet-break time.}
    \label{fig:new3}
\end{figure}

The point at which we see differences in the afterglow behaviour due to the jet structure for an on-beam observer within $\theta_c$ is at $\sim$ the jet-break time. The sharpness of this break can be measured by fitting a simple empirical function for the transition behaviour from the pre- to post-break lightcurve. A parameter, $\kappa$, is used to define the sharpness of the observed break in a smoothly broken power-law as \citep[e.g.][]{beuermann1999, rhoads1999, rossi2004, gorosabel2006, liang2007, schulze2011, vaneerten2013, wang2015, wang2018},
\begin{equation}
    F(t) = \left[F_1^{\kappa}+F_2^{\kappa}\right]^{-1/\kappa},
    \label{eq:sharp}
\end{equation}
where, $F_x = f_b (t/t_b)^{\alpha_x}$ and the subscript `$x$' indicates the trend pre- and post-jet-break, i.e. $\propto t^{-3(p-1)/4}$ and $\propto t^{-p}$ for the case where the emitting frequency is between the characteristic synchrotron frequency, $\nu_m$, and the cooling frequency, $\nu_c$, as $\nu_m<\nu<\nu_c$, and relevant for all our fiducial models, and $f_b$ and $t_b$ indicate the observed flux and time at the jet break.

We fit the model structured jet afterglow lightcurves with the approximation given by equation \ref{eq:sharp} for the behaviour of the lightcurve break.
In fitting equation \ref{eq:sharp} to the model data we fix the earliest (latest) time to $\sim 2 ~t_p(20 ~\tilde{t_b})$, except for the 2C model, where the latest time is fixed at $10 ~\tilde{t_b}$ to avoid significant late-time variability in the post-break lightcurve, here $\tilde{t_b}$ is the expected jet-break time for a TH jet with the same $\theta_c$ and $\iota$.
The best-fit parameters, $[{\alpha_1,~\alpha_2,~\kappa,~f_b,~t_b}]$, for each jet structure are found using a non-linear least-squares minimisation.
For each jet structure model, we evaluate the best-fit value of $\kappa$ for jets described by core sizes $\theta_c=[0.02,~0.04,~0.07,~0.1,~\&~0.2]$, and for each of these at an inclination from $\iota = 0\rightarrow\theta_c$ at step sizes of $0.05\theta_c$.
The value of $\kappa$ and limits ($1\sigma$) are shown in Figure \ref{fig:new2} as a function of the inclination, $\iota/\theta_c$.
The model lightcurves and the equation \ref{eq:sharp} best-fit for each jet structure model at $\iota=0.0$ is shown in Figure \ref{fig:new3}.

We further apply this method to measure the sharpness of the afterglow jet break to a sample of GRB lightcurves.
Using GRBs with multi-band afterglows that are consistent with the closure relations,  the `gold' sample in \cite{wang2015}, we select those with an achromatic jet break listed in \cite{wang2018}.
The optical afterglows for the resulting 20 GRBs are collected and composite lightcurves with the flux at various filters shifted, using the spectral energy distribution (SED) for each burst and corrected for Galactic foreground extinction and host contribution (if necessary and possible), to produce $R_C$-band afterglows at $z=1$ \citep{kann2010, kann2011}.
We use a Markov Chain Monte Carlo (MCMC) via \texttt{emcee} \citep{emcee} to fit the parameters $[{\alpha_1,~\alpha_2,~\kappa,~f_b,~t_b}]$ in equation \ref{eq:sharp} to the observed data around the jet-break time.
The preferred $\kappa$, or sharpness parameter, can then be used as a test for jet structure in these GRB afterglows.
{ Although the GRB afterglows in our sample have achromatic jet breaks at X-ray frequencies, we use only the optical frequency afterglow to measure the sharpness of the break.
This avoids introducing model dependent cooling frequencies, $\nu_c$, whose evolution depends on the medium.
\cite{vaneerten2013} showed that for the case $\nu>\nu_c$, the jet break lightcurve has a higher $\kappa$ value at all inclinations within the jet opening angle than it would have in the case of $\nu<\nu_c$.}

The starting point for each GRB afterglow fit is given by the $\alpha_{1},~\alpha_2$ and $t_b$ values\footnote{For GRBs 050922C, 060729, 080710, and 091127, the jet-break times reported in Table 2 of \cite{wang2018} differ from those in their Fig. 1; in these cases, we use values for $t_b$ found from the figure.} with associated uncertainties reported in \cite{wang2018} and the break time shifted to a source at $z=1$.
The initial guess for $f_b$ is the interpolated lightcurve flux from the data at $t_b$, and $\kappa = 3\pm0.5$ -- for the sharpness parameter, \cite{rhoads1999} favoured a $\kappa=0.4$, however, by fitting $\alpha_{1,2}$ to afterglow data, \cite{liang2007} found $\kappa \sim 3$ was preferred in most cases and consistent with the mean value found by \cite{zeh2006} for seven GRBs where $\kappa$ was freely fit; $\kappa=3.1\pm1.6$.
The prior is flat for each parameter in a range;
$(0\leq\alpha_1\leq3,~\alpha_1\leq\alpha_2\leq4,~0.001\leq\kappa\leq10,~-2\leq\log_{10}{f_b}\leq6)$ and flat for $\log_{10}{t_b}$ for each burst with unique limits set by the data constraints.
We use 1,000 walkers and 15,000 steps per GRB.
To avoid variability in the early afterglow data due to either a reverse shock, energy injection or flares related to the prompt emission, we set an earliest time, $t_{\rm min}$ for each GRB.
In three cases, the long GRBs\,050408 and 130427A and the short GRB\,130603B, we also set a maximum time, beyond which the data shows variability that is inconsistent with the late-time single power-law behaviour (for GRB 130603B, the late-time excess is a kilonova, e.g. \citealt{tanvir2013, berger2013}).

\section{Results}\label{sec:3}

\begin{table*}
\centering
\caption{Best fit absolute parameter ranges over all inclinations for our model afterglows. Values in parentheses are those found for the non-spreading jets.}
\label{table:t2}
\begin{tabular}{c c c c c}
Model & $\theta_c$ & $\Delta\alpha~[{\rm min} - {\rm max}]\pm~{\rm err}$ & $\kappa~[{\rm min} - {\rm max}]\pm~{\rm err}$ & $\theta_{j,{\rm inferred}}[\iota=0.0 - \iota=\theta_c]$ \\
\hline
TH & 0.02 & $[ 1.31 ( 1.05 ) -  1.69 ( 1.36 )]\pm 0.06 ( 0.04 )$ & $[ 2.72 ( 2.23 ) -  3.93 ( 3.51 )]\pm 0.82 ( 0.43 )$ & $[ 0.02 ( 0.02 ) - 0.04 ( 0.04 )]$\\
TH & 0.04 & $[ 1.51 ( 1.20 ) -  1.81 ( 1.46 )]\pm 0.04 ( 0.03 )$ & $[ 1.44 ( 1.11 ) -  3.42 ( 2.68 )]\pm 0.51 ( 0.20 )$ & $[ 0.04 ( 0.04 ) - 0.06 ( 0.07 )]$\\
TH & 0.07 & $[ 1.57 ( 1.31 ) -  1.88 ( 1.71 )]\pm 0.05 ( 0.05 )$ & $[ 0.97 ( 0.54 ) -  3.52 ( 2.25 )]\pm 0.52 ( 0.16 )$ & $[ 0.07 ( 0.07 ) - 0.10 ( 0.12 )]$\\
TH & 0.1 & $[ 1.73 ( 1.44 ) -  1.99 ( 2.97 )]\pm 0.06 ( 0.23 )$ & $[ 0.57 ( 0.14 ) -  3.35 ( 1.99 )]\pm 0.51 ( 0.14 )$ & $[ 0.09 ( 0.10 ) - 0.12 ( 0.18 )]$\\
TH & 0.2 & $[ 1.39 ( 1.29 ) -  1.81 ( 2.66 )]\pm 0.06 ( 0.23 )$ & $[ 0.63 ( 0.15 ) -  3.29 ( 1.64 )]\pm 0.70 ( 0.13 )$ & $[ 0.16 ( 0.20 ) - 0.22 ( 0.33 )]$\\
G& 0.02 & $[ 1.36 ( 1.17 ) -  1.65 ( 1.43 )]\pm 0.05 ( 0.04 )$ & $[ 1.52 ( 1.64 ) -  2.17 ( 2.06 )]\pm 0.28 ( 0.20 )$ & $[ 0.02 ( 0.02 ) - 0.03 ( 0.03 )]$\\
G& 0.04 & $[ 1.57 ( 1.32 ) -  1.81 ( 1.54 )]\pm 0.04 ( 0.03 )$ & $[ 1.24 ( 1.15 ) -  1.62 ( 1.33 )]\pm 0.21 ( 0.12 )$ & $[ 0.04 ( 0.04 ) - 0.06 ( 0.06 )]$\\
G& 0.07 & $[ 1.67 ( 1.43 ) -  1.82 ( 1.57 )]\pm 0.04 ( 0.04 )$ & $[ 0.98 ( 0.78 ) -  1.52 ( 1.09 )]\pm 0.19 ( 0.10 )$ & $[ 0.06 ( 0.07 ) - 0.09 ( 0.11 )]$\\
G& 0.1 & $[ 1.68 ( 1.48 ) -  1.82 ( 1.64 )]\pm 0.04 ( 0.04 )$ & $[ 0.89 ( 0.64 ) -  1.54 ( 1.01 )]\pm 0.20 ( 0.09 )$ & $[ 0.08 ( 0.09 ) - 0.13 ( 0.16 )]$\\
G& 0.2 & $[ 1.56 ( 1.52 ) -  1.75 ( 1.79 )]\pm 0.05 ( 0.06 )$ & $[ 0.89 ( 0.48 ) -  1.62 ( 0.89 )]\pm 0.25 ( 0.08 )$ & $[ 0.14 ( 0.20 ) - 0.21 ( 0.30 )]$\\
2C & 0.02 & $[ 1.35 ( 1.12 ) -  1.79 ( 1.53 )]\pm 0.05 ( 0.03 )$ & $[ 2.25 ( 1.67 ) -  3.24 ( 2.63 )]\pm 0.47 ( 0.22 )$ & $[ 0.02 ( 0.02 ) - 0.04 ( 0.04 )]$\\
2C & 0.04 & $[ 1.23 ( 1.17 ) -  1.90 ( 1.66 )]\pm 0.07 ( 0.05 )$ & $[ 1.37 ( 0.69 ) -  4.21 ( 2.06 )]\pm 1.86 ( 0.20 )$ & $[ 0.04 ( 0.04 ) - 0.06 ( 0.07 )]$\\
2C & 0.07 & $[ 0.89 ( 1.15 ) -  1.80 ( 1.98 )]\pm 0.07 ( 0.14 )$ & $[ 1.56 ( 0.31 ) -  4.60 ( 1.80 )]\pm 1.85 ( 0.17 )$ & $[ 0.07 ( 0.08 ) - 0.08 ( 0.12 )]$\\
2C & 0.1 & $[ 1.04 ( 1.43 ) -  1.62 ( 3.36 )]\pm 0.06 ( 0.80 )$ & $[ 0.77 ( 0.11 ) -  4.58 ( 1.67 )]\pm 1.66 ( 0.13 )$ & $[ 0.09 ( 0.11 ) - 0.12 ( 0.24 )]$\\
2C & 0.2 & $[ 1.46 ( 1.68 ) -  2.10 ( 3.14 )]\pm 0.10 ( 0.35 )$ & $[ 0.33 ( 0.14 ) -  2.62 ( 0.96 )]\pm 0.31 ( 0.10 )$ & $[ 0.16 ( 0.24 ) - 0.24 ( 0.66 )]$\\
PL & 0.02 & $[ 1.25 ( 1.18 ) -  1.54 ( 1.45 )]\pm 0.02 ( 0.02 )$ & $[ 1.44 ( 1.30 ) -  1.83 ( 1.57 )]\pm 0.10 ( 0.09 )$ & $[ 0.03 ( 0.03 ) - 0.04 ( 0.04 )]$\\
PL & 0.04 & $[ 1.42 ( 1.35 ) -  1.63 ( 1.65 )]\pm 0.03 ( 0.05 )$ & $[ 0.71 ( 0.55 ) -  1.51 ( 1.39 )]\pm 0.08 ( 0.08 )$ & $[ 0.05 ( 0.05 ) - 0.07 ( 0.09 )]$\\
PL & 0.07 & $[ 1.44 ( 1.46 ) -  2.15 ( 2.55 )]\pm 0.07 ( 0.19 )$ & $[ 0.32 ( 0.21 ) -  1.16 ( 0.99 )]\pm 0.06 ( 0.08 )$ & $[ 0.08 ( 0.11 ) - 0.14 ( 0.19 )]$\\
PL & 0.1 & $[ 1.72 ( 1.66 ) -  2.72 ( 3.16 )]\pm 0.14 ( 0.33 )$ & $[ 0.23 ( 0.16 ) -  0.68 ( 0.72 )]\pm 0.05 ( 0.07 )$ & $[ 0.13 ( 0.18 ) - 0.20 ( 0.33 )]$\\
PL & 0.2 & $[ 1.54 ( 1.74 ) -  2.03 ( 3.13 )]\pm 0.06 ( 0.38 )$ & $[ 0.67 ( 0.14 ) -  1.22 ( 0.70 )]\pm 0.16 ( 0.08 )$ & $[ 0.23 ( 0.43 ) - 0.24 ( 0.85 )]$\\

\hline
\end{tabular}
\end{table*}

For the model jet afterglows we find $\Delta\alpha=\alpha_1-\alpha_2$, and $\kappa$ ranges for the five core angles and inclinations using the best-fit values of equation \ref{eq:sharp} to the model data.
The analytic estimate for the opening angle of an on-axis, $\iota=0.0$, TH jet is given by \citep[e.g.][]{sari1999, nakar2002},
\begin{equation}
    \theta_{j,{\rm inferred}} \sim 0.055 \left(\frac{n/0.001~{\rm cm}^{-3}}{E/10^{52}~{\rm erg}}\right)^{1/8} \left(\frac{t_b}{{\rm day}}\right)^{3/8}~~~ {\rm radians}
    \label{eq:open}
\end{equation}
where we have used\footnote{The inferred jet opening angles are larger by a factor, $\sim1.3$ when using $t\sim R/(4\Gamma^2 c)$, resulting in typically better agreement for $\theta_c$ from the TH models with lateral spreading where $\theta_c\gtrsim0.1$, and/or for those viewed at $\iota=\theta_c$.} $t\sim R/(2\Gamma^2 c)$.
With this expression we estimate the opening angle as would be inferred from the best-fit jet-break time for each model.
Table \ref{table:t2} lists the ranges found for $\Delta\alpha$ and $\kappa$, and the inferred $\theta_j$ for the cases where $\iota=0.0$ and $\iota = \theta_c$.
Figure \ref{fig:new2} shows the $\kappa$ at each inclination within the jet core angle for the four jet structures.
Within each panel the results for core sizes $0.02\leq\theta_c\leq0.2$ are shown with the discrete core sizes given by a different colour line according to the legend.
For the TH jet structure, the results of \cite{vaneerten2013} are shown with pink lines for the three core sizes that they tested; the observed behaviour is qualitatively the same to that seen with our models.
We include a very narrow jet model in our sample with $\theta_c=0.02$, for this model $\kappa(\iota)$ remains relatively high at all inclinations when compared to models with a core size $>0.02$.

From Table \ref{table:t2}, we note that although $\Delta\alpha$ is consistent between all models, the sharpness of the break, $\kappa$, has a typically lower value for jet structures with a continuously declining energy/Lorentz factor lateral distribution, i.e. the smooth-edged jet structures (G, PL); see also Figure \ref{fig:new2}.
Where a model lightcurve is more variable at either very early or very late times, then fitting equation \ref{eq:sharp} can result in either artificially sharp breaks (high values of $\kappa$) or very soft breaks ($\kappa\ll1$).
This is especially pronounced in the fits for the 2C model, where the wider jet component contributes at late times resulting in a highly variable $\alpha(t)$ post-jet-break (see Figure \ref{fig:new1} and \ref{fig:new3}).
This is reflected in the comparatively noisy curves in Figure \ref{fig:new2} and uncertainties for $\kappa$ and $\Delta\alpha$ outliers in Table \ref{table:t2}.

\begin{figure*}
\includegraphics[width=\textwidth]{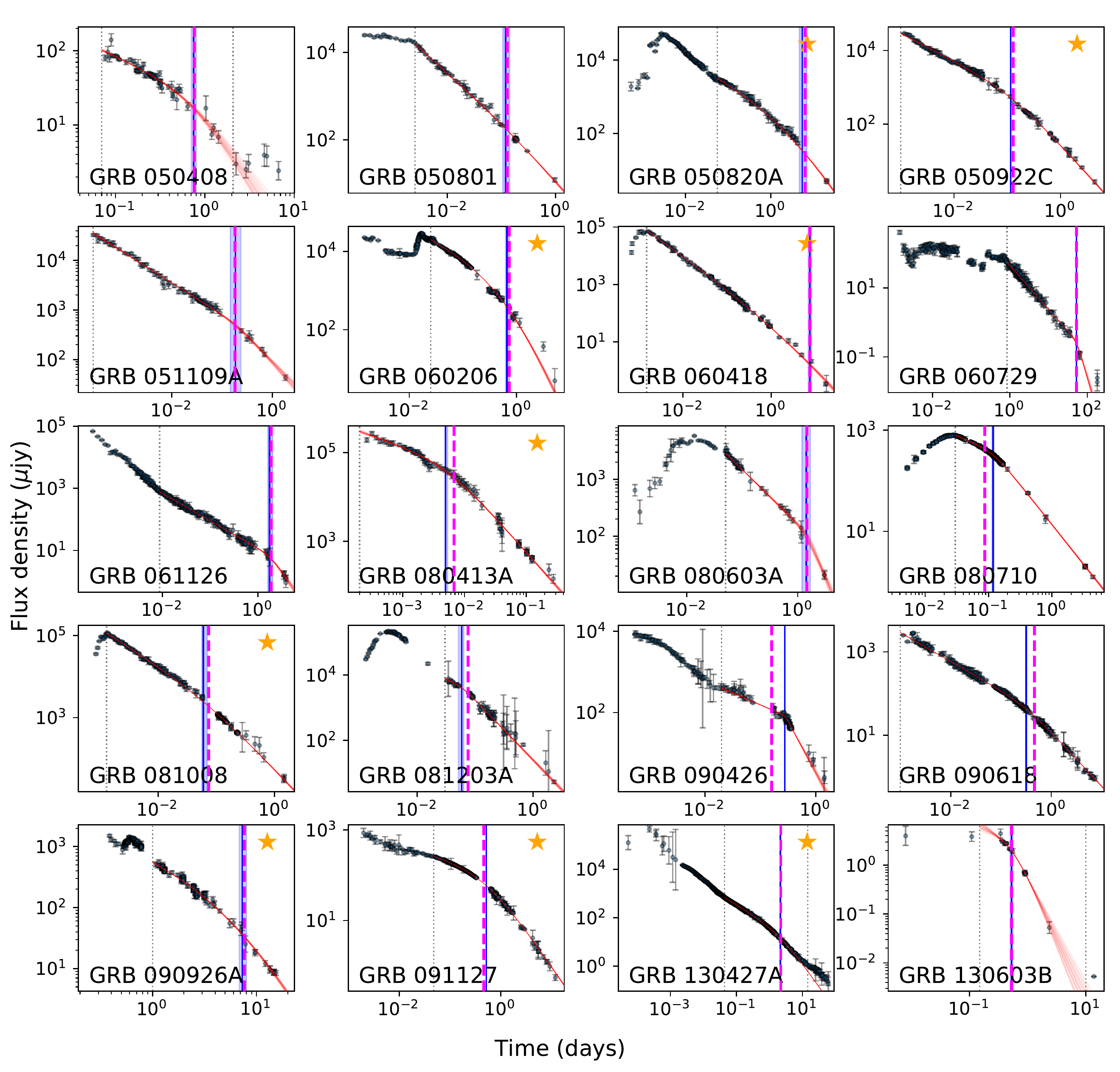}
\caption{The afterglow lightcurves for 20 GRBs that are reported to be consistent with a jet-break scenario. All sources shifted to $z=1$ and data traces the $R_C$-band emission. A sample of 100 parameter sets are randomly drawn from the MCMC posterior for each burst to construct the red-line jet-break approximation. The corresponding break time, $t_b$, is shown as a vertical blue line and the region between the 16th and 84th percentiles is shaded. The pink dashed line indicates the $t_b$ estimate from \protect\cite{wang2018} used as a prior and shifted to $z=1$. Dotted grey lines indicate $t_{\rm min}$, and where required $t_{\rm max}$, for the allowed data range of the fit. Gold stars indicate GRB afterglows that are consistent with being a structured jet viewed from within the core angle, see \S\ref{sec:4}.}
\label{fig:f5}
\end{figure*}

The central parameter values for $\alpha_1,~\alpha_2,~t_b,{\rm and}~\kappa$ from the MCMC fits of equation \ref{eq:sharp} for the 20 GRBs in our sample are listed in Table \ref{table:t1} and the lightcurves shown in Figure \ref{fig:f5}; all lightcurves here are shifted to a redshift $z=1$.
Each panel shows the data for the named GRB, the red curves show a random sample of 100 parameter sets for equation \ref{eq:sharp} from the MCMC fit to the data.
The value of most interest for this study is the best fit value of $\kappa$, the sharpness parameter.
Whereas $\alpha_1$ can depend on the details of the spectral regime, the distribution of shocked electrons, and the nature of the ambient environment; and $\alpha_2$ on the details of the lateral spreading and $p$; the value for the sharpness of the transition, $\kappa$, depends on the lateral structure of the jet, the inclination to the line-of-sight, and the width of the jet.

\begin{table*}
\centering
\caption{Our sample of 20 GRB afterglows that obey the closure relations and literature claims of an achromatic jet break. The MCMC fitted parameters for a lightcurve given by equation \ref{eq:sharp}. The reported times are observer times for a burst at $z=1$. }
\label{table:t1}
\begin{tabular}{c c c c c c c c c}
GRB & $t_{\rm min(max)}$ (days) & $\alpha_1$ & $\alpha_2$ & $t_b$ (days)$^{1}$ & $\kappa$ & $\theta_j$ (rad), medium$^2$ & { Category$^{3}$} & Refs.$^{3}$ \\
\hline
050408  & 0.07(2.07) & $0.56^{+0.06}_{-0.09}$ & $1.92^{+0.17}_{-0.16}$ & $0.75^{+0.05}_{-0.04}$ & $1.28^{+0.60}_{-0.44}$ & 0.12 W & Either & [1][4--9]\\
050801  & 0.0025 & $1.09^{+0.03}_{-0.04}$ & $1.41^{+0.05}_{-0.04}$ & $0.12^{+0.02}_{-0.02}$ & $2.94^{+2.27}_{-1.37}$ & 0.05 I & Either & [1][10--12]\\
050820A & 0.056 & $0.32^{+0.18}_{-0.17}$ & $2.39^{+0.18}_{-0.20}$ & $5.43^{+1.50}_{-0.72}$ & $0.16^{+0.09}_{-0.05}$ & 0.10 I & Smooth & [1][13--14]\\
050922C & 0.001 & $0.75^{+0.01}_{-0.01}$ & $1.60^{+0.02}_{-0.02}$ & $0.116^{+0.002}_{-0.001}$ & $1.39^{+0.16}_{-0.15}$ & 0.03 I & Smooth & [1][15--21]\\
051109A & 0.00027 & $0.64^{+0.01}_{-0.01}$ & $1.10^{+0.08}_{-0.07}$ & $0.19^{+0.05}_{-0.04}$ & $6.84^{+2.19}_{-2.55}$ & 0.06 W & Sharp & [1][22--31]\\
060206  & 0.025 & $0.72^{+0.03}_{-0.03}$ & $3.12^{+0.14}_{-0.08}$ & $0.66^{+0.07}_{-0.03}$ & $0.40^{+0.04}_{-0.04}$ & 0.06 W & Smooth & [1][32--45]\\
060418  & 0.0015 & $1.05^{+0.03}_{-0.04}$ & $2.03^{+0.05}_{-0.05}$ & $7.52^{+0.63}_{-0.57}$ & $0.35^{+0.06}_{-0.06}$ & 0.03 I & Smooth & [1][46--53]\\
060729  & 0.85 & $1.265^{+0.004}_{-0.005}$ & $3.80^{+0.14}_{-0.25}$ & $52.47^{+0.042}_{-0.30}$ & $3.22^{+2.67}_{-0.72}$ & 0.36 W & Sharp & [1][2][54--58]\\
061126  & 0.0087 & $0.88^{+0.01}_{-0.01}$ & $2.18^{+0.25}_{-0.17}$ & $1.76^{+0.26}_{-0.14}$ & $3.66^{+2.00}_{-1.21}$ & 0.08 I & Sharp & [1][59--60]\\
080413A & 0.0002 & $0.50^{+0.06}_{-0.07}$ & $1.57^{+0.06}_{-0.05}$ & $0.005^{+0.000}_{-0.000}$ & $1.30^{+0.46}_{-0.36}$ & 0.02 W & Smooth & [1][61--64]\\
080603A & 0.05 & $0.95^{+0.02}_{-0.02}$ & $2.31^{+0.38}_{-0.26}$ & $1.42^{+0.26}_{-0.23}$ & $5.86^{+2.74}_{-2.50}$ & 0.09 I & Sharp & [2][65--67]\\
080710  & 0.03 & $0.52^{+1.06}_{-0.06}$ & $1.58^{+0.02}_{1.07}$ & $0.118^{+0.003}_{-0.004}$ & $3.32^{+0.85}_{-0.72}$ & 0.06 I & Sharp & [1][68--74] \\
081008  & 0.0013 & $0.88^{+0.03}_{-0.03}$ & $1.52^{+0.05}_{-0.04}$ & $0.060^{+0.010}_{-0.004}$ & $1.36^{+0.37}_{-0.32}$ & 0.04 I & Smooth & [1][85]\\
081203A & 0.03 & $0.80^{+0.13}_{-0.30}$ & $1.84^{+0.03}_{-0.03}$ & $0.059^{+0.004}_{-0.004}$ & $6.20^{+2.62}_{-2.84}$ & 0.03 I & Sharp & [1][76--83]\\
090426  & 0.02 & $0.59^{+0.01}_{-0.01}$ & $2.53^{+0.09}_{-0.09}$ & $0.280^{+0.002}_{-0.003}$ & $9.15^{+0.62}_{-1.10}$ & 0.10 W & Sharp & [84--90]\\
090618  & 0.001 & $0.690^{+0.002}_{-0.002}$ & $1.30^{+0.01}_{-0.01}$ & $0.32^{+0.01}_{-0.01}$ & $2.83^{+0.14}_{-0.14}$ & 0.03 I & Sharp & [91]\\
090926A & 1.00 & $0.48^{+0.26}_{-0.27}$ & $3.35^{+0.27}_{-0.28}$ & $7.29^{+0.74}_{-0.45}$ & $0.29^{+0.17}_{-0.09}$ & 0.07 I & Smooth & [1][2][92]\\
091127  & 0.048 & $0.32^{+0.02}_{-0.02}$ & $1.64^{+0.02}_{-0.02}$ & $0.52^{+0.01}_{-0.01}$ & $0.88^{+0.07}_{-0.07}$ & 0.07 I & Smooth & [3][93--99]\\
130427A & 0.044(14.5) & $0.834^{+0.003}_{-0.003}$ & $1.94^{+0.01}_{-0.01}$ & $2.14^{+0.01}_{-0.01}$ & $1.10^{+0.02}_{-0.02}$ & 0.05 I & Smooth & [3]\\
130603B & 0.15(10.0) & $0.66^{+0.29}_{-0.35}$  & $2.51^{+0.33}_{-0.25}$ & $0.53^{+0.03}_{-0.03}$ & $2.00^{+2.48}_{-1.00}$ & 0.10 I & Either & [96]  \\
\hline
\end{tabular}
\\$^{1}$ Half the observed jet-break time,
corresponding to the source-frame time if the source was at $z=1$.
$^2$ The literature opening angle as listed in \cite{zhao2020} for the preferred medium of our GRBs as stated in \cite{wang2015}, ISM/wind (I/W).
{ $^{3}$ The category that best describes the jet-break shape according to our fits; smooth-, sharped-edged jet structure, or either. }
$^{3}$References for the data used: [1] \cite{kann2010}, [2] \cite{kann2011}, [3] \cite{kann2019}, 
[4] \cite{2005GCN..3200....1W}, [5] \cite{2005GCN..3258....1M}, [6] \cite{2005GCN..3561....1F}, [7] \cite{2005GCN..3261....1K}, [8] \cite{2006ApJ...645..450F}, [9] \cite{2007A&A...462L..57D}, [10] \cite{2009ApJ...702..489R}, [11] \cite{2007MNRAS.377.1638D}, [12] \cite{2005GCN..3728....1M}, [13] \cite{2006ApJ...652..490C}, [14] \cite{2006Natur.442..172V}. [15] \cite{2009ApJ...702..489R}, [16] \cite{2005GCN..4095....1L}, [17] \cite{2005GCN..4018....1O}, [18] \cite{2005GCN..4023....1D}, [19] \cite{2005GCN..4026....1H}, [20] \cite{2005GCN..4027....1N}, [21] \cite{2005GCN..4041....1H}, [22] \cite{2005GCN..4218....1M}, [23] \cite{2005GCN..4220....1H}, [24] \cite{2007ApJ...657..925Y}, [25] \cite{2005GCN..4295....1K}, [26] \cite{2005GCN..4259....1M}, [27] \cite{2005GCN..4240....1L}, [28] \cite{2005GCN..4239....1W}, [29] \cite{2005GCN..4235....1H}, [30] \cite{2005GCN..4230....1H}, [31] \cite{2005GCN..4216....1B}, [32] \cite{2010A&A...523A..70T}, [33] \cite{2007MNRAS.381L..65C}, [34] \cite{2007ApJ...654L..21S}, [35] \cite{2006ApJ...648.1125M}, [36] \cite{2006ApJ...642L..99W}, [37] \cite{2006GCN..4768....1R}, [38] \cite{2006GCN..4750....1L}, [39] \cite{2006GCN..4722....1B}, [40] \cite{2006GCN..4716....1T}, [41] \cite{2006GCN..4702....1A}, [42] \cite{2006GCN..4699....1M}, [43] \cite{2006GCN..4691....1O}, [44] \cite{2006GCN..4696....1L}, [45] \cite{2006GCN..4732....1G}, [46] \cite{2006GCN..4966....1F}, [47] \cite{2006GCN..4971....1N}, [48] \cite{2006GCN..4978....1S}, [49] \cite{2010ApJ...711..641C}, [50] \cite{2008ApJ...686.1209M}, [51] \cite{2007A&A...469L..13M}, [52] \cite{2006GCN..4984....1H}, [53] \cite{2006GCN..4982....1C}, [54] \cite{2010MNRAS.401.2773S}, [55] \cite{2009ApJ...702..489R}, [56] \cite{2007ApJ...662..443G}, [57] \cite{2006GCN..5366....1Q}, [58] \cite{2011MNRAS.413..669C}, [59] \cite{2008ApJ...687..443G}, [60] \cite{2008ApJ...672..449P}, [61] \cite{2008AIPC.1065..103Y}, [62] \cite{2008GCN..7622....1F}, [63] \cite{2008GCN..7597....1A}, [64] \cite{2008GCN..7595....1K}, [65] \cite{2011MNRAS.417.2124G}, [66] \cite{2008GCN..7810....1S}, [67] \cite{2008GCN..7793....1M}, [68] \cite{2009A&A...508..593K}, [69] \cite{2008GCN..7973....1Y}, [70] \cite{2008GCN..7970....1P}, [71] \cite{2008GCN..7967....1W}, [72] \cite{2008GCN..7963....1B}, [73] \cite{2008GCN..7960....1D}, [74] \cite{2008GCN..7959....1L}, [75] \cite{2010ApJ...711..870Y}, [76] \cite{2008GCN..8645....1R}, [77] \cite{2008GCN..8618....1L}, [78] \cite{2008GCN..8617....1W}, [79] \cite{2008GCN..8604....1V}, [80] \cite{2008GCN..8695....1F}, [81] \cite{2008GCN..8629....1I}, [82] \cite{2008GCN..8619....1M}, [83] \cite{2008GCN..8615....1A}, [84] \cite{2011A&A...531L...6N}, [85] \cite{2011MNRAS.414..479T}, [86] \cite{2011MNRAS.410...27X}, [87] \cite{2009A&A...507L..45A}, [88] \cite{2009GCN..9292....1K}, [89] \cite{2009GCN..9285....1M}, [90] \cite{2009GCN..9267....1Y}, [91] \cite{2011ApJ...732...29C}, [92] \cite{2010ApJ...720..862R}, [93] \cite{2011A&A...535A.127V}, [94] \cite{2011A&A...535A..57F}, [95] \cite{2010ApJ...718L.150C}, [96] \cite{2014A&A...563A..62D}

\end{table*}

\begin{figure*}
    \centering
    \includegraphics[width=\textwidth]{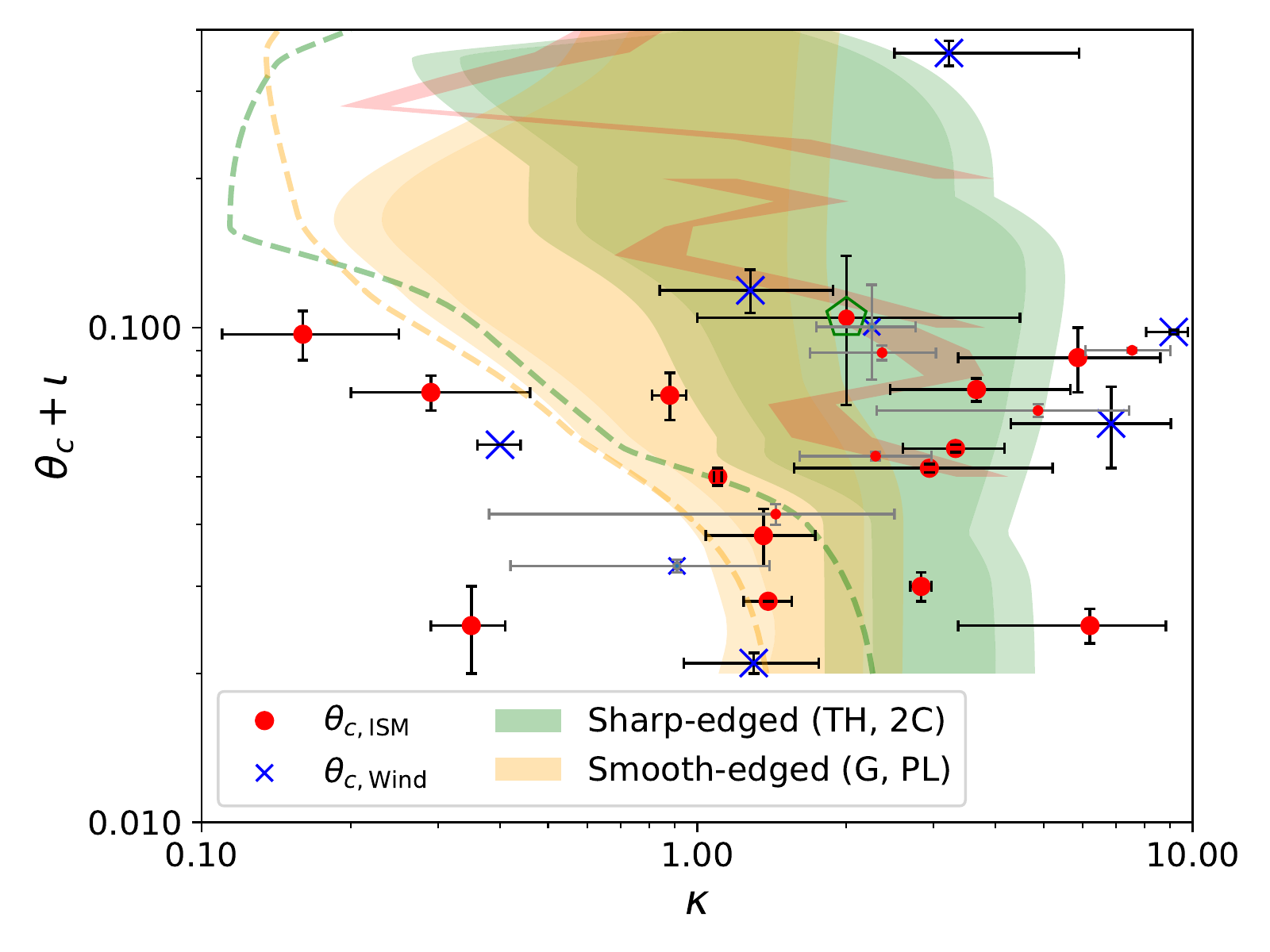}
    \caption{
    The jet break sharpness parameter $\kappa$ vs.~the opening-angle plus inclination, $\theta_c+\iota\equiv \theta_{j,{\rm inferred}}$, for our model GRB afterglows. The parameter region indicative of the sharp-edged structured jets (TH, 2C) with spreading is shown with green shading; the parameter region for smooth-edged jets (G, PL) with spreading is shown with yellow shading. A 20\% uncertainty is added to the edges to give an indication for a change in $p$ within our fiducial models.
    For non-spreading jets, the parameter regime at $\kappa\lesssim2$ is indistinguishable where $\theta_c+\iota\gtrsim0.05$ -- dashed lines indicate the low-$\kappa$ limits for the non-spreading models. The red shaded regions indicate the values of $\kappa$ from \protect\cite{vaneerten2013} for three TH jets using hydrodynamical simulations. The best values for the inferred jet opening angles, $ \theta_{j,{\rm inferred}}$, and their preferred medium (ISM -- red circle, wind -- blue cross), are shown \protect\citep{zhao2020} for the bursts in our sample (black error bars) and those from \protect\cite{zeh2006} (smaller markers and grey error bars). The short GRB 130603B is indicated by the green pentagon.}
    \label{fig:results}
\end{figure*}

\section{Discussion}\label{sec:4}

We have used model afterglows from four distinct jet structures, two with sharp-edged cores (TH and 2C), and two with smooth-edged cores (G and PL), to show how the shape of the jet break for an observer aligned within the core opening angle can provide clues to the presence of lateral jet structure.
We fit these afterglows with a commonly used smoothly-broken power-law function leaving $\kappa$, the parameter that describes the sharpness of the break, as a free parameter.
Figure \ref{fig:new2} shows $\kappa<2.2$ for smooth-edged jet structures, whereas sharp-edged jet structures show a wider range that depends on the inclination of the system and the width of the jet core, resulting in $\kappa\lesssim5$.

For sharp-edged structures, e.g. TH and 2C, as an observer's line-of-sight to the jet central axis increases within the core opening angle then the value of $\kappa$ typically dips before increasing again as $\iota\sim\theta_c$; with the exception of very narrow jet cores, $\theta_c\sim0.02$.
Whereas, for smooth-edged jet structures, the value of $\kappa$ within the jet core remains $\sim$constant or mildly increasing with inclination.
From Figure \ref{fig:new1}, we can see that for the TH and 2C jet structures, where $\iota=0.0$ or $\iota=\theta_c$, the temporal index $\alpha$ is $\sim$constant before the sharp change at the jet break, however, between these inclination limits, a quasi-break caused by the near edge of the jet core becoming visible can be seen.
This quasi-break results in a small change in $\alpha$ before the break and when fit by equation \ref{eq:sharp}, a lower value of $\kappa$ for these inclinations.
For comparison, the values of $\kappa$ for three jet widths from \cite{vaneerten2013} are shown in the top-left panel of Figure \ref{fig:new2} as pink dashed/dotted lines.
\cite{vaneerten2013} found that the range of $\kappa$ values is larger for wider jets with a minimum $\kappa$ at $\sim$mid-way between the central axis and the jet edge, our results confirm this where $\theta_c\geq0.04$.
For the G and PL jet structures, the smooth edge of the jet core softens or eliminates this quasi-break resulting in a more continuously declining $\alpha$ before the jet-break time and when fit by equation \ref{eq:sharp}, a value for $\kappa$ that is typically $\kappa\lesssim2$ at all inclinations.

For jet models where lateral spreading at the sound speed is included, the fit value for $\kappa$ is typically higher than where no lateral spreading is included and contrary to the findings of \cite{rossi2004}\footnote{As noted in \S \ref{sec:2} and in \cite{rossi2004}, the time-interval over which the lightcurve is fit can affect the returned fit parameters. We have carefully chosen our time-interval for the model lightcurves so that the returned fits of equation \ref{eq:sharp}, for most cases, closely follows the model lightcurve during the jet-break transition.}.
For a laterally spreading outflow, the jet break transition results in a larger $\Delta\alpha$, see Table \ref{table:t2}. 
Where the timescale of the jet break is comparable between the spreading and non-spreading models for the jet-break, a larger $\Delta\alpha$ will result in a larger $\kappa$.
Similarly, for `mid-way' inclinations where a quasi-break is present, then lateral spreading results in a more pronounced feature that is responsible for the larger variability in $\kappa$ with $\iota$ for the same $\theta_c$ models.

As equation \ref{eq:sharp} is a function that asymptotically approaches a single power-law at either extreme, then for cases where early- and/or late-time variability in the model lightcurves is present, some care must be taken when windowing the lightcurve for the fit.
We set maximum and minimum times for each structure in an attempt to avoid the worst variability, however, this can result in time windows that are too narrow, and the resulting fit does not accurately reproduce the expected behaviour.
Examples of this can be seen in Table \ref{table:t2} for e.g. the 2C and the PL models with $\theta_c=0.1$ and $0.2$, where some cases have $\Delta\alpha\gtrsim2.75$; here the change in $\alpha$ across the jet break should fall in the range $0.75\leq\Delta\alpha\leq2.75$, where $2\leq p \leq3$, and the lower limit is set by the edge effect \citep{zhang2006, gao2013}.
An example of such a poor fit is apparent in Figure \ref{fig:new3}, showing $\iota=0.0$ lightcurves where the inferred break-time for the non-spreading PL model with $\theta_c=0.2$ is much later than expected (see $\theta_{j,{\rm inferred}}=0.43$, and more than twice the expected $0.2$, in Table \ref{table:t2}).
The fit lightcurve here is cut off before the single power-law behaviour is observed resulting in the higher than expected $\Delta\alpha$ for this example.
These cases highlight how small amounts of variability at early/late times, and/or short time windows can skew the fit results for a smoothly-broken power-law function.
When fitting such a function to observed data to determine the best-fit value of $\kappa$, these factors should be carefully considered as the resulting fit parameter can easily occupy high values, i.e. $\kappa\gg10$, that are difficult to reconcile with theoretical expectations. 

The post-break temporal behaviour for the afterglow from our models, as shown in Figure \ref{fig:new1}, is beyond the simple theoretical approximations of $\alpha_1-3/4$ and $-p$ for the non-spreading and spreading cases respectively \citep{sari1999}.
This is a known result \citep[e.g.][]{granot2007} and the steep decline for the top-hat case with lateral spreading is consistent with hydrodynamic results \citep[e.g.][]{zhang2009, decolle2012, vaneerten2013}.
At optical to X-ray frequencies, these hydrodynamic simulations find that the steepest temporal index post jet-break is $\alpha\lesssim-3.1$ (where $p=2.5$) and consistent with the $\alpha\sim-3.3$ shown in Figure \ref{fig:new1} for our model, based on the maximal spreading at the sound speed.
A post jet-break temporal index that is steeper than the analytical expectation for both a spreading and non-spreading jet is also found by \cite{lu2020}.
Although the expansion description that we use can be considered an upper limit, the resulting temporal behaviour of the afterglow lightcurve at about the jet-break time is consistent with that seen in full, and computationally expensive, hydrodynamic simulations.

From Figure \ref{fig:new3} and Table \ref{table:t2}, we can see that where lateral spreading is included, the jet break occurs at a marginally earlier time than for the non-spreading case for most examples -- where we expect the peak, or deceleration time, for the spreading/non-spreading cases to be the same.
Using the fitted $t_b$, which may over- or underestimate the real model break time (see discussion above), for each model with equation \ref{eq:open}, an estimate of $\theta_c+\iota$  can be found.
Although this analytic estimate is based on the break-time for a TH jet, we find a reasonable estimate for all jet structure models with $\theta_c\lesssim0.07$, and best for the cases with $\iota=0.0$ and without lateral spreading.
This analytic estimate is based on the timescale for the $\Gamma(t)=\theta_c^{-1}$, and so we expect the best results from cases that are most similar to this, i.e. non-laterally spreading TH jets.
This estimate assumes that lateral spreading does not affect the evolution of the lightcurves until after the jet-break time; however, the shell will undergo some lateral expansion at all times during the deceleration.
At early times, where this expansion is mild and confined to the edges of the jet, as energy is conserved, then the radial extent of the blastwave will gradually fall behind that of an identical but non-spreading outflow; the radial extent of the blastwave determines the observed timescale as ${\rm d}t_{\rm obs}/{\rm d}R = 1/(\beta(R) c) - \cos(\iota)/c$, and thus, for a jet with significant lateral spreading, the jet-break time will be marginally sooner than the case without.
This is more apparent for wider-cored jet structures, where the timescale to the jet break is longer and spreading is more significant.
A similar effect can be seen in \cite{lu2020} where they compare lightcurves with and without lateral expansion, and in \cite{granot2007} where the lightcurves from hydrodynamic codes and non-lateral spreading semi-analytic lightcurves are shown.

We have fit a smoothly-broken power-law estimate for the afterglow flux behaviour, equation \ref{eq:sharp}, to the observed optical afterglows of 20 GRBs (including one short GRB).
The data and a sample of the lightcurves from an MCMC posterior distribution for each burst are shown in Figure \ref{fig:f5}.
The best-fit parameters for the temporal indices, $\alpha_1$ and $\alpha_2$, the sharpness parameter, $\kappa$, and the observed jet-break time, $t_b$, assuming a source at redshift $z=1$, are listed in Table \ref{table:t1}.
The jet-break time, $t_b$, is equivalent to the time when the entire jet is within the beaming cone, $1/\Gamma$, and for a mildly inclined jet, the opening angle estimate found from $t_b$ will be equivalent to $\theta_c+\iota$, where $\iota\leq\theta_c$.
Assuming a top-hat jet structure for all GRBs should, therefore, result in a strong inverse correlation for $\kappa\lesssim2$ with the jet opening angle, i.e. lower $\kappa$ having larger $\theta_c+\iota\simeq\theta_{j,{\rm inferred}}$.

In Figure \ref{fig:results} we show  the $\kappa ~{\rm vs}~ \theta_c+\iota$ for our GRB sample using the opening angles from \citet[][and references therein]{zhao2020} and the preferred environment (uniform ISM or wind) from the closure relation fits by \cite{wang2015} to highlight the most likely $\theta_c+\iota$ value.
Additionally, we show, using a smaller marker size and grey error bar lines, the seven GRBs with fitted $\kappa$ from \cite{zeh2006}; GRBs 990510, 000301C, 010222A, 020813, 030226, 030329, 041006 -- opening angles are from \cite{zhao2020} except for GRB 041006, where we use the value in \cite{zeh2006}.
We also show the expected parameter space from our models with a green shaded region for sharp-edged jet structures (TH and 2C) and with a yellow shaded region for a smooth-edged jet structure (G and PL).
The limits on the parameter space are those for a jet with lateral spreading, and the lower limit on $\kappa$ for the non-spreading jets is indicated with a dashed line for each case respectively.
Additionally, we show the values of $\kappa$ from the hydrodynamic simulations for TH jets from \cite{vaneerten2013}, where $2\leq p\leq3$ and $\theta_c=[0.05,~0.01,~0.2]+\iota$ as a pink shaded region; note that a lower $p$ results in a higher value for the sharpness $\kappa$.
We see no correlation between $\kappa$ and increasing $\theta_c$ for our sample.

Half of the GRBs in our sample have $\kappa<2.0$, and are potentially consistent with the expectation for a smooth-edged structured jet (G and/or PL).
By considering the effect of viewing angle and jet width on the sharpness, this number drops to nine GRBs:
GRBs 050820A, 050922C, 060206, 060418, 080413A, 081008, 090926A, 091127, and 130427A -- each marked with a gold star in Figure \ref{fig:f5}.
An additional two GRBs from the seven in \cite{zeh2006} fit our criteria:
GRBs 010222A and 020813.
In some cases, the data at or after the jet-break time are sparse -- in these cases, e.g. GRBs 050820A and 060418, the shape of the pre-break lightcurve and the jet-break time from multi-band fits \citep{wang2018} can help to determine the sharpness of the break, however, as with the non-spreading $\theta_c=0.2$ model PL structure case discussed previously, the poorly sampled lightcurve after the jet break can result in an artificially low $\kappa$ value and may be the cause of the $\kappa=0.16$ and $0.35$ for these two GRBs, respectively.
Similarly, GRB 090926A, with $\kappa=0.29$, has gaps in the data on either side of the jet break; and GRB 060206 with $\kappa=0.40$ has some variability and significant data gaps on either side of the break.
The remaining five GRBs in our candidate structured jet list are well-sampled on either side of the break, and in some cases, throughout, e.g. GRB 050922C, 080413A, and 130427A.
For these better-sampled lightcurves, the sharpness of the break and the inferred opening angles are consistent with smooth-edged jet structures with lateral spreading, i.e. $0.88\lesssim\kappa\lesssim1.39$.

We find three GRBs that have inconclusive values for $\kappa$ in terms of jet structure -- GRBs 050408, 050801 and 130603B.
The remaining eight GRBs are all consistent with being candidate sharp-edged jet structures with $\kappa>2$, of these there are five with afterglow lightcurves well-sampled on either side of the break and two throughout the jet break, e.g. GRBs 080710 and 090618.
Of these five: 
GRB 051109A has sparse data before and throughout the break;
GRB 081203A has sparse data pre-break, where we have discounted very early data as being most likely a reverse-shock component, { or a pre-wind termination evolution};
and, finally, GRB 090426 has a well-measured jet break sharpness of $\kappa=9.15^{+0.62}_{-1.1}$, and is therefore inconsistent with all of the models tested here.
We propose that energy injection, flaring, or a difficult data reduction could be responsible for the apparently very sharp jet-break shape.

Where data are absent at jet-break time, the sharpness of the break derived from the fits should be treated with caution.
High cadence, multi-band (particularly at optical frequencies, where the spectral regime is expected to be consistent, i.e. $\nu<\nu_c$) are essential in determining the temporal behaviour of the afterglow before, during, and after the jet-break time.
For GRBs with $\kappa<2$ and $\theta_c+\iota\gtrsim0.05$, the break sharpness cannot reliably identify candidate smooth-edged jet structure afterglows.
However, where $\theta_c+\iota<0.05$ and $\kappa<2$, then we do not expect afterglows from sharp-edged jet structures to fit this parameter space and such GRBs should be considered as candidate smooth-edged jet structure examples.
Similarly, for $\kappa\gtrsim3$ and all $\theta_c+\iota$ we do not expect any afterglows from smooth-edged jet structures, and such structure profiles can be ruled out in these cases.
A well-sampled afterglow lightcurve that is found to be consistent with a smooth-edged jet structure via this jet-break shape analysis should be followed-up with detailed lightcurve and burst modelling to determine the best fitting jet structure profile \citep[e.g.][]{cunningham2020}.

The single short GRB in our sample sits well within the sharp-edged parameter space, however, it is on the the upper boundary for a smooth-edged structure profile, and so we cannot rule this out.
We note that, although the short GRB 170817A, viewed at $\iota>\theta_c$, is commonly fit with a smooth-edged jet structure profile, i.e. a G or PL model, the afterglow can also be fit with a sharp-edged jet structure, e.g. model A in \cite{lamb2019a}, the energy injection scenario in \cite{lamb2020}, the hydrodynamic results for a TH jet from \cite{gill2019}, a magneto-hydrodynamic, or a hollow-cored jet \cite{nathanail2021}. Thus if we assume that all short GRBs have a universal structure model, then analysis of the shape of the jet break for short GRBs observed within the jet opening angle could be used to indicate the preferred jet structure profile -- sharp-edged vs smooth-edged. 

Within Figure \ref{fig:new2} we show the typical inclination for GRBs within $\theta_c$ as a vertical dashed grey line at $0.57\theta_c$, the value found by \cite{ryan2015} for a GRB afterglow sample.
Where the opening angle of the jet is found from the observed jet-break time using the usual analytic relation \cite[e.g.][]{sari1999, frail2001, nakar2002}, then considering the typical inclination within the jet, $\theta_c \approx 0.6~\theta_{j,{\rm inferred}}$.
The range for the sharpness parameter that we find at $\iota=0.57\theta_c$ is:
for sharp edge structured jets, $0.1\lesssim\kappa\lesssim4.0$;
for smooth edge structured jets, $0.2\lesssim\kappa\lesssim2.0$ -- we note that these values are for model afterglows with $p=2.5$ and the jet structure profiles for our fiducial models.
\cite{vaneerten2013} showed that the change in $\kappa$ due to different $p$ is $\sim\pm10\%$ of the $\kappa(p=2.5)$ value for $2\leq p\leq3$ with lower $p$ resulting in higher $\kappa$, or sharper jet breaks.
In either jet-structure case, the typical $\iota=0.57\theta_c$ value of $\kappa$ is inversely proportional to the jet opening angle.
For very narrow jets, $\theta_c\lesssim0.03$, then $\kappa$ for sharp-edged vs smooth-edged structure profiles is distinct.

Our fiducial models have assumed a uniform ambient density, however, the environment of a GRB may be described by a wind model where the density, $n\propto R^{-k}$ with $k=2$ for a stellar wind \citep[e.g.][]{dai1998, chevalier1999}.
{ Pre-GRB mass injections other than winds e.g., stellar pulsations or common-envelope events for progenitors in a binary, could also affect the properties of the surrounding interstellar medium. 
We do not attempt to model their impact on the GRB afterglow.
}
The nature of the medium, principally, affects the temporal index, $\alpha_1$ \citep[e.g.][]{granot2002, zhang2006, gao2013}. 
{ For a wind, as $k\rightarrow2$, the change in temporal index, $\Delta\alpha$, is reduced when compared to the $\Delta\alpha$ within a uniform medium \citep{granot2007, decolle2012}.
However, \cite{decolle2012} show that the break time is earlier for higher $k$, and despite the smoother change in $\alpha$, the sharpness (or steepening) of the lightcurve across the jet break occurs over a significantly shorter duration. 
We expect competing contributions to the sharpness of the jet break, $\kappa$, with the smaller $\Delta\alpha$ through the jet break within a wind environment resulting in a lower $\kappa$ value\footnote{{ Analogous to that seen in the comparison between outflows with and without lateral spreading.}}, whereas the reduced timescale of the
break would lead to a higher $\kappa$ value.
The two effects should largely balance out at the observable precision to give a similar sharpness measure for jet breaks within a wind medium to those seen for a uniform medium.
As noted by \cite{decolle2012}, the inclination within the jet opening angle has a more pronounced effect on the shape of the jet break than the steepness of the external density profile.
Additionally, we note that the preferred medium for the GRBs in our sample is shown in Fig. \ref{fig:results} with a blue cross for wind and red dot for uniform medium, where we note no preference for wind-medium GRBs in $\kappa$ distribution when compared to the uniform medium GRBs in our sample.
}

{ A jet that is initially propagating within a wind medium may, at late times, transition into a uniform medium as the outflow passes the wind termination shock \citep{chevalier2004}; a statistical study of GRB environments favours a typically small wind termination radius \citep{schulze2011}, meaning that for most jets, the jet break will occur within a uniform medium.
As the afterglow shock passes through the wind termination shock, a change in the temporal index of the afterglow is expected i.e. $(3p-1)/4$ vs $3(p-1)/4$ where $\nu_m<\nu<\nu_c$ and $p>2$ for wind $k=2$ vs uniform medium, with the evidence of such passage in some GRB afterglows \citep[e.g.][]{gendre2007, kamble2007, jin2009, fraija2017, li2020, li_2_2021} and an absence of any wind in others \citep[e.g.][]{bardho2016}.}
Similarly, energy injection can change the afterglow decline \citep[e.g.][]{zhang2006};
where the injection is continuous, then the effect on the jet-break shape is not expected to be significant.
However, where the energy injection is discrete, then variability in the afterglow decline is expected \citep[e.g.][]{kumar2000, zhang2006, laskar2015, lamb2019b}, this variability could skew a smoothly broken power-law function fit if it occurs at early or late times, as discussed above, alternatively, if the discrete energy injection episode coincides with the jet break, then the fit $\kappa$ will be affected.
We expect such a situation to be rare, however, where this does occur, then the inferred value of $\kappa$ is likely to be larger for a discrete energy injection episode.
The large $\kappa=9.15^{+0.62}_{-1.10}$ for GRB 090426, and the apparent variability at the jet-break time in the afterglow lightcurve (see Figure \ref{fig:f5}), could be an example of such an energy injection.
Where the injected energy has a stratified profile \citep[e.g.][]{rees1998, sari2000, nakamura2006, cheng2021} as in Model 2 in \cite{lamb2020}, and the injection timescale coincides with the jet-break time, then the shape of the break will be softened, resulting in a low $\kappa$ value.
For very narrow jets, $\theta_c\lesssim0.04$, this could enable $\kappa\lesssim1$, although, any information on the potential angular jet structure profile will be lost.

\section{Conclusions}\label{sec:5}

We have shown that for an observer at a line-of-sight inclination angle $\iota$ within the core opening angle $\theta_c$ of a GRB, the temporal evolution of the afterglow lightcurve around the jet-break time is distinct between core-dominated jets with a sharp-edged core and those with a smooth-edged core profile.
Using a smoothly-broken power-law function as a way to measure the jet-break sharpness, we show that:
\begin{itemize}
    \item Jets with lateral spreading typically have higher values of the sharpness parameter, $\kappa$, than identical jets without lateral spreading.
    \item Smooth-edged jet structures have a sharpness parameter $\kappa \lesssim 2.2$ with little dependence on the inclination of the system within $\theta_c$.
    \item Sharp-edged jet structures have a more diverse range, $0.1\lesssim\kappa\lesssim4.6$, where $\kappa$ depends strongly on $\theta_c$ and the inclination.
    
\end{itemize}


Using a sample of GRB afterglows (19 long GRBs, one short GRB) that are consistent with an achromatic jet break and the standard closure relations { \citep{wang2015}}, we find that approximately half are candidates for a smooth-, and half for a sharp-edged jet structure profile. 
We find:
\begin{itemize}
    \item Nine candidate smooth-edged jet structure GRBs, including three that have well-sampled lightcurves through the jet break.
    \item Eight candidate sharp-edged jet structure GRBs, including two with well-sampled lightcurves.
    \item Three ambiguous GRBs, consistent with either a smooth- or a sharp-edged jet structure.
\end{itemize}
Of the nine candidate smooth-edged jet structure GRBs, we find five that have well-sampled lightcurves on either side of the break, with three being well-sampled throughout the jet break -- GRBs 081008, 091127 and GRBs 050922C, 080413A, and 130427A as the three well-sampled examples.
Of the eight candidate sharp-edged jet structure GRBs, we find five that are sufficiently well-sampled on either side of the jet break, with two being continuously sampled throughout -- GRBs 051109A, 081203A, and 090426, and GRBs 080710 and 090618 as the two well-sampled examples.
For the remaining four (three) GRBs that sit in the smooth- (sharp-)edged jet structure candidate group, we consider the afterglow lightcurve data to be either too sparse, variable, or to contain significant gaps that could bias the estimated $\kappa$ parameter.

We encourage high-cadence multi-filter optical observations around the jet-break time, for GRBs with a bright optical afterglow.
Understanding the fraction of GRB jets that are consistent with lateral jet structure is important for off-axis, orphan afterglow searches and rates \citep[e.g.][]{ghirlanda2015, lamb2018a, huang2020};
additionally, the fraction of short GRBs showing evidence of jet structure is important for gravitational-wave counterpart search strategies and counterpart rates \citep[e.g.][]{lamb2017, coughlin2020, gompertz2020}.
The presence and diversity of lateral structure within the afterglows to GRBs can also help constrain the physics that drives and shapes these high-energy transients \citep[e.g.][]{pescalli2015, salafia2015, salafia2020, beniamini2019, gottlieb2020, gottlieb2021, hamidani2021, nathanail2021, takahashi2021}.

\section*{Acknowledgements}
We thank the anonymous referee for constructive comments that have improved the work.
GPL and JJF thank Shiho Kobayashi for useful discussions and comments on a pre-submission version.
GPL thanks Lekshmi Resmi for valuable discussions.
GPL is supported by the STFC via grant ST/S000453/1. DAK acknowledges support from Spanish National Research Project RTI2018-098104-J-I00 (GRBPhot), and thanks S. Schulze and M. Bla\v{z}ek for calculations support.
IM is a recipient of the Australian Research Council Future Fellowship FT190100574.

\section*{Data Availability}
Data available on request.



\bibliographystyle{mnras}
\bibliography{ms} 




\appendix

\bsp	
\label{lastpage}
\end{document}